\documentstyle[12pt]{article}
\newcommand{\z}{&&\hspace*{-1cm}}
\begin{document}

\setcounter{page}{0}
\thispagestyle{empty}

{}~\vspace{2cm}

\begin{center} {\Large {\bf
The Gegenbauer Polynomial Technique:\\ the evaluation of a class of
Feynman
diagrams}} \\
 \vspace{1.5cm}
 {\large
  A.V.Kotikov\footnote{On leave of absence from Particle Physics
    Laboratory, JINR, Dubna, Russia.\\ e-mail:
    KOTIKOV@LAPPHP0.IN2P3.FR; KOTIKOV@SUNSE.JINR.DUBNA.SU}\\
 \vspace {0.5cm}
 {\em Laboratoire de Physique Theorique ENSLAPP\\
LAPP, B.P. 110, F-74941, Annecy-le-Vieux Cedex, France}}
\end{center}

\vspace{4.5cm}\noindent
\begin{center} {\bf Abstract} \end{center}

We extend Gegenbauer Polynomials technique
to evaluate a class of complicated Feynman diagrams. New results in the form
of $_3F_2$-hypergeometrical series of unit argument, are presented.
%Some examples of applications of these results are given
As a by-product, we present a new transformation rule for
$_3F_2$-hypergeometric series with  argument $-1$.

%PACS numbers:

\newpage
\pagestyle{plain}

%\section{Introduction}

Though Quantum Field Theory
%(QFT)
 already has a long history and a
number of different approaches, Feynman diagrams (FD) are still the
main source of
its dynamical information.
%the information abour its dynamics.
%and the
The necessity to know
more exactly about the characteristics of physical processes and
specific QFT functions themselves stimulates the calculation of
radiative corrections of ever higher order. This, in its turn, leads
to
%the creation of
new methods of calculation (see, for review,
Ref.\cite{1}). Fifteen years ago
%Chetyrkin, Kataev and Tkachov presented (see \cite{2})
the method based on the expansion of propagators in Gegenbauer series
(see \cite{1.1}) has been introduced in \cite{2,2.1}.
 One has shown  \cite{2} that by this
method the analytical evaluation of counterterms in the minimal
subtraction scheme at the 5-loop level in any model and for any
composite operator was indeed possible. The Gegenbauer Polynomial (GP)
 technique has been applied
successfully for propagator-type FD in many calculations (see \cite{1}-
\cite{2.2}).
In the present paper we present {\it a new
%some
development of the GP technique and apply
it to the evaluation of a class  of master
two-loop diagrams containing the vertex with two propagators having
the powers 1 or $D/2-1$ in $D$-dimensional space}.

%\section{Basic Formulae}

Thoughout the paper we use the following notation. The use of dimensional
regularization is assumed. All the calculations are performed in the
coordinate space\footnote{Note that the choice of $x$-space
   is not crucial. If the reader prefers to work in
  $p$-space he can consider the variable $\{x_i,y_i, ...\}$ as some
  momenta in $p$-space.}
 of dimension $D=4-2\varepsilon$.
%Note that in contrast to the momentum space, we integrate here over
%all internal vertices rather than over the loops.
Any diagram is converted from
$p$-space to $x$-space by applying the Fourier transform
(marked by the symbol $ \stackrel{f}{\Longrightarrow}$)
$$ \frac{1}{(p^2)^{\nu}}  \stackrel{f}{\Longrightarrow}
\int \frac{d^Dp \exp(ipx)}{(p^2)^{\nu}}~=~\pi^{D/2}2^{D-2\nu}
\frac{a_0(\nu)}{(x^2)^{D/2-\nu}},
{}~~~~a_n(\nu)~=~\frac{\Gamma (D/2-\nu +n )}{\Gamma(\nu)},$$
\noindent ($\Gamma$ is the Euler $\Gamma$-function) or by considering  the
dual diagram (see, for example, \cite{3}). The trasformation
%of an initial diagram
to the dual diagram will be denoted by
%is marked by the symbol
  $\stackrel{d}{=}$.

 The dual diagram is obtained from the original one by
%the substitution $p_i \longleftrightarrow x_i$
replacing $p_i$ by $x_i$ into  corresponding integral and
% accordance with
using the rules of
correspondence between the graph and the integral, as in the
%with the corresponding Feynman rules in
$x$-space.
Thus, after applying the Fourier transform we obtain a different
$D$-space integral but the $x$-space
FD similar graphically to corresponding $p$-space one. Considering the
dual transformation, we do not change the integrand, except for the
replacement $p_i \to x_i$, but obtain another graphical
representation.
Using of the both transformations leads to define the number of
really independent FD (see section {\bf 2}).

Note that contrary to \cite{2} we
analyse FD directly in  $x$-space which allows
us to avoid the appearence of Bessel functions.
It is possible in the case of propagator-type FD, because their
dependence on a single external momentum (in $p$-space) or on a single
external coordinate is power-like and known beforehand. The point of
interest is the coefficient function $C_f$, which depends on
$D=4-2\varepsilon$ and is a Laurent series in $\varepsilon $. If
some diagram has $l$-lines\footnote{
 With the lines of the graphs there are associated power-law factors of
 the type $(x^2)^{-\alpha}$, in which $\alpha$ is called the line
 index.}
 with the corresponding indices $\alpha_i$,
$i= \{1,...,l\}$ and $L$-loops in $p$-space, then  its coefficient
functions in $p$-space and
$x$-space are connected by
$$
C_f(p-space)= \frac{\prod_{i=1}^l a_0(\alpha_i)}{a_0(\sum_{i=1}^l
  \alpha_i - DL/2)} \cdot C_f(x-space) ~~{\mbox and }~~
C_f(p-space)=C_f(x-space)
$$
in the case of the Fourier and dual transform, respectively.

%With the lines of the graphs there are associated power-law factors of
%the type $(x^2)^{-\alpha}$, in which $\alpha$ is called the line
%index; to two arrows with index $n$ there corresponds $n$ vectors
%$x^{\mu_1}...x^{\mu_n}$. If the index appears in braskets $(n)$, this
%means a traceless product (TP) of $n$ vectors: $x^{(\mu_1...\mu_n)}$
%(see definition in Appendix B). The
%index of a line is equal to 1 are not marked.
%%The line with index 0
%%and $n$ (traceless) vectors is marked by dotted line.
%When the points $x$
%and $y$ are accompanied by the signs $''+''$ and $''-''$, respectively, this
%means $ \Theta (x^2-y^2)$ in integrand. Here we present the
%following examples:
%$$ \frac{(y-x)^{\mu_1...\mu_n}}{(x-y)^{2\alpha}}~\equiv~$$ %%
%
%$$ \frac{(y-x)^{\mu_1}...(y-x)^{\mu_n}}{(x-y)^{2\alpha}}
%(z-x)^{\mu_1...\mu_n}\Theta(x^2-z^2)~\equiv~$$

Following \cite{2,3} we introduce the traceless product (TP)
$x^{\mu_1...\mu_n}$ connected with the usual product
$x^{\mu_1}...x^{\mu_n}$ by the following equations (see \cite{2,3})
\begin{eqnarray}
\z x^{\mu_1...\mu_n}~=~\hat{S}
\sum_{p \geq 0} \frac{n!(-1)^{p} \Gamma(n-p+\lambda)}
{2^{2p} p! (n-2p)!\Gamma(n+\lambda)}~
g^{\mu_1\mu_2}...g^{\mu_{2p-1}\mu_{2p}}~x^{2p}~x^{\mu_{2p+1}}...x^{\mu_n}
%\label{B1}   \\
               \nonumber  \\
\z x^{\mu_1}...x^{\mu_n}~=~\hat{S}
\sum_{p \geq 0} \frac{n! \Gamma(n-2p+\lambda+1)}
{(2)^{2p} p! (n-2p)!\Gamma(n-p+\lambda+1)}~
g^{\mu_1\mu_2}...g^{\mu_{2p-1}\mu_{2p}}~x^{2p}~x^{\mu_{2p+1}...\mu_n},
\label{B1}
\end{eqnarray}
where hereafter $\lambda \equiv D/2-1$.

Comparing Eqs.(\ref{A4})
%, (\ref{A6})
and (\ref{B1}), we obtain the
following relation between TP and GP
\begin{eqnarray}
z^{\mu_1...\mu_n}~x^{\mu_1...\mu_n}~=~ \frac{n! \Gamma(\lambda)}
{2^n \Gamma(n+\lambda)}~C_n^{\lambda}(\hat{x}\hat{z})~{(x^2 z^2)}^{n/2}
  \label{B2}
\end{eqnarray}
We give also the simple but quite useful conditions:
\begin{eqnarray}
z^{\mu_1...\mu_n}~x^{\mu_1...\mu_n}~=~
z^{\mu_1}...z^{\mu_n}~x^{\mu_1...\mu_n}~=~
z^{\mu_1...\mu_n}x^{\mu_1}...x^{\mu_n},
  \label{B3}
\end{eqnarray}
which
 follow immediately from the TP definition:
$g^{\mu_i\mu_j}~x^{\mu_1...\mu_i...\mu_j...\mu_n}~=~ 0$.

The use of the TP $x^{\mu_1...\mu_n}$ makes it possible to ignore
terms of the type $g^{\mu_i\mu_j}$ that arise upon integration: they can
be readily recovered from the general structure of the TP. Therefore,
in the process of integration it is only necessary to follow the
coefficient of the leading term $x^{\mu_1}...x^{\mu_n}$ (see also
section 3).
%By differeting $\frac{\partial}{\partial x_{\mu}}$ n-times
%to the chain we have the following rule:
The rules to integrate FD containing TP can be found, for example in
\cite{3,3.1}. For a chain we have (hereafter $Dx \equiv (d^Dx)/(2\pi)^D$):
\begin{eqnarray}
%\z
%~~
\int Dx \frac{ x^{\mu_1...\mu_n}}{x^{2\alpha}(x-y)^{2\beta}}
{}~=~
\frac{1}{(4\pi)^{D/2}}~
\frac{y^{\mu_1...\mu_n}}{y^{2(\alpha + \beta - \lambda -1)}}
{}~ A^{n,0}(\alpha, \beta),
%\nonumber \\ \z
%~ \stackrel{(\beta =  \lambda)}{=}~
% \frac{1}{(4\pi)^{D/2}}~\frac{y^{\mu_1...\mu_n}}{y^{2(\alpha -1)}}
%~\frac{1}{\Gamma(\lambda)}
%\frac{1}{(\alpha -1)(n+\lambda +1-\alpha)},
  \label{4}
\end{eqnarray}\\
where
$$A^{n,m}(\alpha,\beta)~=~
\frac{a_n(\alpha)a_m(\beta)}{a_{n+m}(\alpha+\beta-\lambda-1)}$$

Note that in our analysis it is necessary to consider more
complicate cases of integration, when the integrand  contains
$\Theta$ functions. Indeed, using the Eqs.(\ref{A2}) and (\ref{B2}),
we can represent
the propagator $(x_1-x_2)^{-2\lambda}$ into the following form:
\begin{eqnarray}
\frac{1}{(x_1-x_2)^{2\lambda}} ~=~
\sum^{\infty}_{n=0} ~\frac{2^n \Gamma(n+\lambda)}{n!\Gamma(\lambda)}
{}~x_1^{\mu_1...\mu_n}~x_2^{\mu_1...\mu_n}~
\Biggl[
\frac{1}{x_2^{2(\lambda +n)}} \Theta(x_2^2 - x_1^2)
{}~+~
\Bigl(x_1^2 \longleftrightarrow x_2^2  \Bigr)  \Biggr]
  \label{B6}
\end{eqnarray}
%This equation was already applied in Section 3.

Using the GP properties from  Appendix A and the connection (\ref{B2})
between GP and TP,
%and the equations of Appendix C for evaluating  Feynman integrals
%with GP in them integrands,
we obtain the  rules for calculating
FD with the $\Theta$-terms and TP.\\

%$A$. The cotribution of a simple loop with vectors on the lines (and
% $\Theta$-functions) is the ordinary product:
%
%\begin{eqnarray}
%~~~=~~~~~
%%\equiv
%%\frac{(y-x)^{\mu_1}...(y-x)^{\mu_n}(y-x)^{\nu_1}...(y-x)^{\nu_m}}
%%{(y-x)^{2(\alpha + \beta)}} \label{1}
%\end{eqnarray}\\
%
%$B$.

{\bf 1.} {\it The rules}
%to integrate} of chains containing the
%$\Theta$-term and
%TP in them integrands can be easily  write using the
%corresponding rules (\ref{C1}) - (\ref{C15}) in the case of GP  and the
%relation (\ref{B2})
%between GP and TP.
%
%The rules
have the following form:
\begin{eqnarray}
\z
%~~~~~\equiv ~
\int Dx \frac{x^{\mu_1...\mu_n}}{x^{2\alpha}(x-y)^{2\beta}}
\Theta(x^2-y^2)
%\nonumber \\ \z
{}~=~
\frac{1}{(4\pi)^{D/2}}~
\frac{y^{\mu_1...\mu_n}}{y^{2(\alpha + \beta - \lambda -1)}}
{}~ \sum_{m=0}^{\infty}
%B(m,n|\beta,\lambda)~
%\frac{\Gamma(m+\beta +n)}{m! \Gamma(m+n+1+\lambda) \Gamma(\beta)}~
%\frac{\Gamma(m+\beta - \lambda)}{\Gamma(\beta - \lambda)}~
\frac{B(m,n|\beta,\lambda)}{m+\alpha + \beta - \lambda -1}~ \nonumber \\
\z \stackrel{(\beta =
  \lambda)}{=}~
 \frac{1}{(4\pi)^{D/2}}~\frac{y^{\mu_1...\mu_n}}{y^{2(\alpha  -1)}}
{}~\frac{1}{\Gamma(\lambda)}
\frac{1}{(\alpha -1)(n+\lambda)}  \label{2}
\end{eqnarray}

\begin{eqnarray}
\z
%~~~\equiv ~
\int Dx \frac{ x^{\mu_1...\mu_n}}{x^{2\alpha}(x-y)^{2\beta}}
\Theta(y^2-x^2)
% \nonumber \\ \z
 ~=~
\frac{1}{(4\pi)^{D/2}}~
\frac{y^{\mu_1...\mu_n}}{y^{2(\alpha + \beta - \lambda -1)}}
{}~ \sum_{m=0}^{\infty}
%B(m,n|\beta,\lambda) ~
%\frac{\Gamma(m+\beta +n)}{m! \Gamma(m+n+1+\lambda) \Gamma(\beta)}~
%\frac{\Gamma(m+\beta - \lambda)}{\Gamma(\beta - \lambda)}~
\frac{B(m,n|\beta,\lambda)}{m+n- \alpha + \lambda +1}~  \nonumber \\
\z \stackrel{(\beta =
  \lambda)}{=}~
 \frac{1}{(4\pi)^{D/2}}~\frac{y^{\mu_1...\mu_n}}{y^{2(\alpha -1)}}
{}~\frac{1}{\Gamma(\lambda)}
\frac{1}{(n+\lambda +1- \alpha )(n+\lambda)}  \label{3}
\end{eqnarray}\\
where
$$B(m,n|\beta,\lambda)~=~\frac{\Gamma(m+\beta +n)}
{m! \Gamma(m+n+1+\lambda) \Gamma(\beta)}~
\frac{\Gamma(m+\beta - \lambda)}{\Gamma(\beta - \lambda)}~~
%\mbox{ and }~~ \lambda=D/2-1
$$

Note that  the $x$-space index of an ordinary line is $\lambda$,
where all our formulae are simplified.
Hereafter
we add  this specific case $\beta = \lambda$ to our rules.
%when the index $\beta$of the expanded propagator equals $\lambda$.

The sum of above diagrams does not contain $\Theta$-terms
%by Eq.(\ref{C5}) for $_3F_2$-hypergeometrical functions (with unit argument)
%can be leaded to the good known form
and should reproduce Eq.(\ref{4}).
%good known results (see, for example \cite{3}):
%\frac{\Gamma(\lambda +1- \alpha +n) \Gamma(\lambda +1- \beta +m)
%\Gamma(\alpha + \beta - \lambda -1)}{\Gamma(\alpha) \Gamma( \beta)
%\Gamma(2 \lambda +2- \alpha - \beta +n+m)}$$
To compare the r.h.s. of Eqs.(\ref{2},\ref{3}) and the r.h.s. of
Eq.(\ref{4}) we use the
transformation of $_3F_2$-hypergeometric function with unit argument
(see \cite{5}):
\begin{eqnarray}
\z
_3F_2(a,b,c;e,f;1)~=~\frac{\Gamma(1-a)\Gamma(e)\Gamma(f)\Gamma(c-b)}
{\Gamma(e-b)\Gamma(f-b)\Gamma(1+b-a)\Gamma(c)} \cdot~
\nonumber  \\
\z
_3F_2(b,b-e+1,b-f+1;1+b-c,1+b-a;1) ~+~ \Bigl(b  \longleftrightarrow c \Bigr)
  \label{C5}
\end{eqnarray}

When $e=b+1$, the $_3F_2$-function can be represented as the sum of another
$_3F_2$-function
%with $e \to b-a+1$ and $a \to b-a$
and a term
containing only $\Gamma$-function products:
%\begin{eqnarray}
%\z
%_3F_2(a,b,c;b+1,f;1)~=~\frac{\Gamma(1-a)\Gamma(b+1)\Gamma(f)\Gamma(c-b)}
%{\Gamma(f-b)\Gamma(1+b-a)\Gamma(c)} \nonumber  \\
%\z
%+~ \frac{b}{b-c} \frac{\Gamma(1-a)\Gamma(f)}{\Gamma(f-c)\Gamma(1+c-a)}
%\cdot~
%_3F_2(c,c-b,c-f+1;1+c-b,1+c-a;1)
%%~+~ \Bigl(b  \longleftrightarrow c \Bigr)
%  \label{8.1}
%\end{eqnarray}
%or
\begin{eqnarray}
\z
 \sum_{k=0}^{\infty} \frac{\Gamma(k+a)\Gamma(k+c)}
{k!\Gamma(k+f)} \frac{1}{k+b}
{}~=~\frac{\Gamma (a)\Gamma(1-a)\Gamma(b)\Gamma(c-b)}
{\Gamma(f-b)\Gamma(1+b-a)} \nonumber  \\
\z
-~  \frac{\Gamma(1-a)\Gamma(a)}{\Gamma(f-c)\Gamma(1+c-f)}
\cdot~
 \sum_{k=0}^{\infty} \frac{\Gamma(k+c-f+1)\Gamma(k+c)}
{k!\Gamma(k+1+c-a)} \frac{1}{k+c-b}
  \label{8.2}
\end{eqnarray}
This is the case (when $k=m, b=\alpha+\beta-\lambda-1 , c=n+\beta$) to
compare Eq.(\ref{4}) and the sum of eqs.(\ref{2},\ref{3}).

Analogously to Eqs.(\ref{2}) and (\ref{3}) we have more
complicate cases\footnote{The full set of rules will be presented in
  \cite{4}.}:
\begin{eqnarray}
\z
%~~~\equiv ~
\int Dx \frac { x^{\mu_1...\mu_n}}{x^{2\alpha}(x-y)^{2\beta}}
\Theta(x^2-z^2)
%%\nonumber \\\z
%=~ \frac{1}{(4\pi)^{D/2}}~
%%\frac{1}{\Gamma(\beta)}~
%y^{\mu_1...\mu_n}~
% \sum_{m=0}^{\infty} ~B(m,n|\beta,\lambda)
%%\frac{\Gamma(m+\beta +n)}{m! \Gamma(m+n+1+\lambda) }~
%%\frac{\Gamma(m+\beta - \lambda)}{\Gamma(\beta - \lambda)} ~
%\nonumber \\
%\z
%~\cdot \Biggl[
%%[
%\frac{1}{y^{2(\alpha + \beta - \lambda -1)}}
%\biggl(
%\frac{1}{m+\alpha + \beta - \lambda -1}~+~
%\frac{1}{m-\alpha +n+1+ \lambda } \biggr) \Theta (y^2-z^2)   \nonumber \\
%\z ~+~  \frac{1}{z^{2(\alpha + \beta - \lambda -1)}}
%\biggl( {\Bigl(\frac{y^2}{z^2}\Bigr)}^m
%\frac{\Theta (z^2-y^2)}{m+\alpha + \beta - \lambda -1}~-~
%{\Bigl(\frac{z^2}{y^2}\Bigr)}^{m+ \beta +n}
%\frac{\Theta (y^2-z^2)}{m-\alpha +n+1+ \lambda } \biggr)
%%]
%\Biggr]   \nonumber \\ \z
      ~ =~ \frac{1}{(4\pi)^{D/2}}~
%\frac{
y^{\mu_1...\mu_n}
%}{\Gamma(\beta)}~
%~[
\Biggr[
%\Big[   A^{n,0}(\alpha,\beta)
\frac{\Theta (y^2 -z^2)}{y^{2(\alpha + \beta - \lambda -1)}}
{}~ A^{n,0}(\alpha,\beta)
 \nonumber \\ \z
+~
 \sum_{m=0}^{\infty}
{}~\frac{B(m,n|\beta,\lambda)}{z^{2(\alpha + \beta - \lambda -1)}}
%\frac{\Gamma (m+\beta +n)}{m! \Gamma (m+n+1+\lambda)\Gamma (\beta ) }~
%\frac{\Gamma(m+\beta - \lambda)}{\Gamma(\beta - \lambda)}~
%  \nonumber \\    \z
%\frac{\Theta (z^2 -y^2)}{z^{2(\alpha + \beta - \lambda -1)}}
 ~ \biggl( {\Bigl(\frac{y^2}{z^2}\Bigr)}^{m}
\frac{\Theta (z^2 -y^2)}{m+\alpha + \beta - \lambda -1}~-~
{\Bigl(\frac{z^2}{y^2}\Bigr)}^{m+ \beta +n}
\frac{\Theta (y^2 -z^2)}{m-\alpha +n+1+ \lambda } \biggr)
%]
 \Biggr]
                         ~ \nonumber \\
\z  \stackrel{(\beta =
  \lambda)}{=}~
 \frac{1}{(4\pi)^{D/2}}~ \frac{1}{\Gamma(\lambda)}~y^{\mu_1...\mu_n}~
%[
\Biggl[
\frac{1}{y^{2(\alpha -1)}}
\frac{\Theta (y^2-z^2)}{(\alpha -1)(n+\lambda +1 -\alpha)}  \nonumber \\
\z ~+~ \frac{1}{z^{2(\alpha -1)}} \frac{1}{n+\lambda }
\biggl(
\frac{\Theta (z^2-y^2)}{\alpha -1}~-~
{\Bigl(\frac{z^2}{y^2}\Bigr)}^{n+ \lambda }
\frac{\Theta (y^2-z^2)}{n+1+ \lambda -\alpha } \biggr)
%]
 \Biggr]  \label{5}
\end{eqnarray}

\begin{eqnarray}
\z
%~~~\equiv ~
\int Dx \frac { x^{\mu_1...\mu_n}}{x^{2\alpha}(x-y)^{2\beta}}
\Theta(z^2-y^2)
%%\nonumber \\  \z
%=~
%\frac{1}{(4\pi)^{D/2}}~\frac{1}{\Gamma(\beta)}
%~y^{\mu_1...\mu_n}~
% \sum_{m=0}^{\infty} ~B(m,n|\beta,\lambda)
%%\frac{\Gamma(m+\beta +n)}{m! \Gamma(m+n+1+\lambda) }~
%%\frac{\Gamma(m+\beta - \lambda)}{\Gamma(\beta - \lambda)}~
%\nonumber \\
%\z
%%[
%~\cdot \Biggl[
%\frac{1}{y^{2(\alpha+\beta-\lambda-1)}}
%\biggl(
%\frac{1}{m+\alpha + \beta - \lambda -1}~+~
%\frac{1}{m-\alpha +n+1+ \lambda } \biggr) \Theta (z^2-y^2)  \nonumber \\
%\z  ~-~ \frac{1}{z^{2(\alpha+\beta-\lambda-1}}
%\biggl( {\Bigl(\frac{y^2}{z^2}\Bigr)}^m
%\frac{\Theta (z^2-y^2)}{m+\alpha + \beta - \lambda -1}~-~
%{\Bigl(\frac{z^2}{y^2}\Bigr)}^{m+ \beta +n}
%\frac{\Theta (y^2-z^2)}{m-\alpha +n+1+ \lambda } \biggr)
%%]
%\Biggr]  \nonumber \\\z
{}~ =~ \frac{1}{(4\pi)^{D/2}}~
%\frac{
y^{\mu_1...\mu_n}~
%}{\Gamma(\beta)}~
%[
\Biggl[
%A^{n,0}(\alpha,\beta)
\frac{\Theta (z^2 -y^2)}{y^{2(\alpha + \beta - \lambda -1)}}
{}~A^{n,0}(\alpha,\beta)
 \nonumber \\ \z
-~   \sum_{m=0}^{\infty}
{}~\frac{B(m,n|\beta,\lambda)}{z^{2(\alpha + \beta - \lambda -1)}}
%\frac{\Gamma (m+\beta +n)}{m! \Gamma (m+n+1+\lambda)\Gamma (\beta ) }~
%\frac{\Gamma(m+\beta - \lambda)}{\Gamma(\beta - \lambda)}~
%\nonumber \\ \z
%\frac{\Theta (z^2 -y^2)}{z^{2(\alpha + \beta - \lambda -1)}}
{}~\biggl( {\Bigl(\frac{y^2}{z^2}\Bigr)}^{m}
 \frac{\Theta (z^2 -y^2)}{m+\alpha + \beta - \lambda -1}~-~
{\Bigl(\frac{z^2}{y^2}\Bigr)}^{m+ \beta +n}
\frac{\Theta (y^2 -z^2)}{m-\alpha +n+1+ \lambda } \biggr)
%]
 \Biggr]
{}~ \nonumber \\
\z  \stackrel{(\beta =
  \lambda)}{=}~
 \frac{1}{(4\pi)^{D/2}}~ \frac{1}{\Gamma(\lambda)}~y^{\mu_1...\mu_n}~
%[
\Biggl[
\frac{1}{y^{2(\alpha -1)}}
\frac{\Theta (z^2-y^2)}{(\alpha -1)(n+\lambda +1 -\alpha)}  \nonumber \\
\z  ~-~ \frac{1}{z^{2(\alpha -1)}} \frac{1}{n+\lambda }
\biggl(
\frac{\Theta (z^2-y^2)}{\alpha -1}~-~
{\Bigl(\frac{z^2}{y^2}\Bigr)}^{n+ \lambda }
\frac{\Theta (y^2-z^2)}{n+1+ \lambda -\alpha } \biggr)
 \Biggr]  \label{6}
\end{eqnarray}\\
One can easily see that the sum of the above diagrams lead to
results identical to (\ref{4}).\\

%\section{The evaluation of a class of Feynman diagrams}

 {\bf 2.} {\it The
aim of this article} is to study  a class of master
two-loop diagrams containing the vertex with two propagators having index
1 or $\lambda$.

%{\bf 1.}
Consider in the $x$-space the following general
diagram
$$ \int  \frac {Dx Dy}{y^{2\alpha}(z-y)^{2t}(z-x)^{2\beta}x^{2\gamma}
(x-y)^{2s}}
{}~\equiv ~J(\alpha,t, \beta, \gamma, s)$$
and restrict ourselves to the FD
$A(\alpha, \beta, \gamma ) ~=~ J(\alpha,\lambda, \beta, \gamma,
\lambda)$,
which is the one
%\footnote{more correctly the CF $C_f[A]$ of
%  $A(\alpha,\beta, \gamma)$.}
of FD of interest for us here. It is easily
shown\footnote{For example, we can consider new FD which is the result
  of gluing of $A(\alpha,\beta, \gamma)$ and the propagator with
index $\sigma + \Delta$, where $  \sigma = 3+ \lambda -(\alpha + \beta
+ \gamma )$ and $\Delta \to 0$. This new FD has a pole $1/\Delta $
with the corresponding coefficient independent from the external
moment. Hence, this coefficient does not depend from
%fact
which
%^from
propagator
of new FD is external and, thus, the values of these different
(nongluing) FD are connected.} that
\begin{eqnarray}
C_f[A(\alpha,\beta, \gamma)]~=~C_f[A(\alpha,\sigma, \gamma)]~=~
C_f[J(\gamma,\lambda, \lambda,\sigma, \alpha)]~=~
C_f[J(\sigma,\gamma,\lambda, \lambda,\beta)],   \label{7}
\end{eqnarray}
%{\bf 2.}
 Doing Fourier transformation of both: the diagram $A(\alpha,\beta,\gamma)$
%from eq.(\ref{8})
and its solution
in the form $C_f[A(\alpha, \beta, \gamma)](z^2)^{-\tilde \sigma}$,
where hereafter
$\tilde t = \lambda +1-t,~t= \{\alpha,\beta, \gamma, \sigma , ...\}$, and
considering the new
diagram as one in the $x$-space (i.e. making the dual transformation)
we obtain the relation
\begin{eqnarray} \z
C_f[A(\alpha,\beta, \gamma)]~ \stackrel{f}{=}
\frac{a_0^2(\lambda)a_0(\alpha)a_0(\beta)a_0(\gamma)}{a_0(\delta)}~
C_f[J(\tilde \alpha,1,\tilde \beta,\tilde \gamma ,1)] \nonumber\\ \z
{}~\stackrel{d}{=}
a_0^2(\lambda)a_0(\alpha)a_0(\beta)a_0(\gamma)a_0(\sigma)~
C_f[J(\tilde \beta,1,\tilde \alpha,\tilde \gamma,1)]~,  \label{9}
\end{eqnarray}
between the diagram, which contains the vertex
with two propagators
% propagators, which applying onevertex and
having the index $\lambda$, with the similar diagram containg the
propagators  with the index 1.

%3.
Repeating the manipulations of subsection 1 we can obtain the
following relations:
\begin{eqnarray}
%\z
C_f[J(\tilde \beta,1,\tilde \alpha,\tilde \gamma,1)]=
C_f[J(\tilde \beta,1,\tilde \sigma,\tilde \gamma,1)]
%\nonumber \\ \z
=C_f[J(1,1,\tilde \gamma ,\tilde \sigma,\tilde \alpha )]=
C_f[J(\tilde \sigma,1,1,\tilde \gamma,\tilde \beta)]  \label{10}
\end{eqnarray}
Thus, we have obtained the relations between all diagrams from the class
introduced in the beginning of this section. Hence, it is necessary
to find the solution for one of them. We prefer to analyse the
diagram $A(\alpha,\beta,\gamma)$, that is the content of the next section.\\

%\section{The calculations}
{\bf 3.}
{\it We calculate the diagram} $A(\alpha, \beta, \gamma)$ by the
following way\footnote{The symbol $\stackrel{(n)}{=}$ markes the fact
  that the equation $(n)$ is used on this step.}:
\begin{eqnarray}
\z A(\alpha, \beta, \gamma)~ \stackrel{(\ref{B6})}{=}~
% \mbox{ from Eq.(\ref{B6}) }=~
\sum_{n=0}^{\infty} \frac{2^n \Gamma(n+\lambda)}{n! \Gamma(\lambda)}
\int Dx Dy \frac { z^{\mu_1...\mu_n}}{x^{2\gamma}(z-x)^{2\beta}}
 \frac { y^{\mu_1...\mu_n}}{y^{2\alpha}(x-y)^{2\lambda}}
\Bigl[\frac{\Theta(z^2-y^2)}{z^{2(n+\lambda)}}~+~
\frac{\Theta(y^2-z^2)}{y^{2(n+\lambda)}} \Bigr]~
%\mbox{ from eq.(\ref{5},\ref{6}) }=~
 \nonumber \\
%\z \frac{1}{(4\pi)^{D/2}}~ \frac{1}{\Gamma(\lambda)}
%\sum_{n=0}^{\infty} \frac{2^n \Gamma(n+\lambda)}{n! \Gamma(\lambda)}
%   \nonumber \\
%\z
%\Biggl[ \frac{1}{\alpha + \lambda +n-1} \frac{1}{1- \alpha}~~~~~~~~+
%\frac{1}{\alpha + \lambda +n-1} \frac{1}{n+ \lambda}~~~~~~~~-
%\frac{1}{n+ \lambda} \frac{1}{1- \alpha}~~~~~~~~  \nonumber \\
%\z  ~+~ \frac{1}{\lambda +n+1-\alpha} \frac{1}{\alpha -1}~~~~~~~~-
%\frac{1}{\lambda +n} \frac{1}{\alpha -1}~~~~~~~~+
%\frac{1}{\lambda +n+1-\alpha} \frac{1}{n+\lambda}~~~~~~~~\Biggr]  \nonumber \\
\z \stackrel{(\ref{5},\ref{6})}{=}~
 \frac{1}{(4\pi)^{D/2}}~ \frac{1}{\Gamma(\lambda)}~\frac{1}{\alpha -1}
\sum_{n=0}^{\infty} \frac{2^n \Gamma(n+\lambda)}{n! \Gamma(\lambda)}
\int Dx \frac {z^{\mu_1...\mu_n}x^{\mu_1...\mu_n} }{x^{2\gamma}(z-x)^{2\beta}}
\cdot \Biggl[
\frac{1}{\lambda +n+1-\alpha}  \nonumber \\
\z \cdot \bigl(
\frac{\Theta(z^2-x^2)}{z^{2(n+\lambda)}x^{2(\alpha -1)}}~+~
\frac{\Theta(x^2-z^2)}{x^{2(n+\lambda)}z^{2(\alpha -1)}} \bigr)
%\nonumber \\  \z
{}~-~
\frac{1}{\lambda +n+\alpha -1} \cdot \bigl(
\frac{\Theta(z^2-x^2)}{z^{2(n+\lambda +\alpha -1)}}~+~
\frac{\Theta(x^2-z^2)}{x^{2(n+\lambda +\alpha -1) }} \bigr)
\Biggr]  \label{11}
\end{eqnarray}
%where
%$$K(n,\lambda)~=~\frac{2^n \Gamma(n+\lambda)}{n! \Gamma(\lambda)}
%$$
%is the coefficient between GP and TP (see Eq.(\ref{B2})).

Evaluating the second integral and using the equation
(see (\ref{A1}) and (\ref{B2}))
$$x^{\mu_1...\mu_n}x^{\mu_1...\mu_n}~=~
%K^{-1}(n,\lambda)
\frac{\Gamma(n+2 \lambda)\Gamma(\lambda
  )}{2^n\Gamma(2\lambda)\Gamma(n+\lambda )},$$
we have from Eqs.(\ref{2}) and (\ref{3}):
\begin{eqnarray}
\z C_f[A(\alpha, \beta, \gamma)]~=~
 \frac{1}{(4\pi)^{D}}~ \frac{1}{\Gamma(\lambda)}~\frac{1}{\alpha -1}
%                            \nonumber \\
%\z
 \sum_{n=0}^{\infty}
\frac{\Gamma(n+2 \lambda )
%a_n(2\lambda)
}{n! \Gamma (2 \lambda) }~
 \sum_{m=0}^{\infty} ~B(m,n|\beta,\lambda)
%\frac{\Gamma (m+\beta +n)}{m! \Gamma (m+n+1+\lambda)\Gamma (\beta ) }~
%\frac{\Gamma(m+\beta - \lambda)}{\Gamma(\beta - \lambda)}~
 \nonumber \\
\z
\cdot \Biggr[
%\Bigl[
\frac{1}{\lambda +n+1-\alpha} \bigl(
\frac{1}{m+n+2-\alpha- \gamma + \lambda} +
\frac{1}{m+n+ \gamma + \beta -1} \bigr)  \nonumber \\
\z ~-~  \frac{1}{\lambda +n+\alpha -1} \bigl(
\frac{1}{m+n+ \lambda +1- \gamma} +
\frac{1}{m+n+\alpha +\gamma + \beta -2} \bigr)
\Biggr]  \label{12}
%\Bigr]
\end{eqnarray}

The r.h.s. of Eq.(\ref{12}) is
%quite good result
already in an easy-to-use form,
because the sums converge  quite
fastly and thus they can be easily evaluated numerically. However,
the aim of our investigations is to obtain the simplest analytical
expresion for the initial diagram. As we will see in this section it is
%the one-time series (
$_3F_2$-hypergeometric functions with unit argument.

To our purposes it is useful to transform the parts
$$ \bigl(
\frac{1}{m+n+2-\alpha- \gamma + \lambda} +
\frac{1}{m+n+ \gamma + \beta -1} \bigr)~~\to~~
 \bigl(
\frac{1}{m+\alpha+ \gamma + \beta- \lambda -2} +
\frac{1}{m+1- \gamma } \bigr)$$

$$\bigl(
\frac{1}{m+n+ \lambda +1- \gamma} +
\frac{1}{m+n+\alpha +\gamma + \beta -2} \bigr)~~\to~~
\bigl(
\frac{1}{m+ \gamma +\beta -1- \lambda} +
\frac{1}{m+2- \alpha -\gamma } \bigr)$$

It is can be done by using of Eq.(\ref{8.2}) or by representating $\Theta(x_i^2
- x_j^2)$ terms into FD of the r.h.s. of Eq.(\ref{11}) as $1-\Theta(x_j^2-
x_i^2)$. Then the result for
$ C_f[A(\alpha, \beta, \gamma)]$ transforms to
$$ C_f[A(\alpha, \beta, \gamma)]~=~ \frac{1}{(4\pi)^{D}}~
\frac{1}{\Gamma(\lambda)}~
\frac{1}{\alpha -1}~
\Bigl[
\overline I ~-~ \tilde I
\Bigr],$$
where
%$$
\begin{eqnarray}
\z \overline{I}~=~
\sum_{n=0}^{\infty}
\frac{\Gamma(n+2 \lambda )
%a_n(2\lambda)
}{n! \Gamma (2 \lambda) }~
\Biggl[
\frac{1}{\lambda +n+1-\alpha} \cdot \biggl(
A^{n,0}(\alpha -1+\gamma, \beta )+A^{n,0}(n+\lambda +\gamma, \beta )
\biggr)
     \nonumber \\
\z ~-~ \frac{1}{\lambda +n+\alpha -1} \cdot \biggl(
A^{n,0}(\gamma ,\beta )+A^{n,0}(n+\alpha + \lambda +\gamma -1, \beta)
\biggr) \Biggr]  \label{13}
\end{eqnarray}
%$$
and
%$$
\begin{eqnarray}
\z
\tilde{I}~=~
%\z
\sum_{n=0}^{\infty}
\frac{\Gamma(n+2 \lambda )
%a_n(2\lambda)
}{n! \Gamma (2 \lambda) }~
 \sum_{m=0}^{\infty} ~B(m,n|\beta,\lambda)
%\frac{\Gamma (m+\beta +n)}{m! \Gamma (m+n+1+\lambda)\Gamma (\beta ) }~
%\frac{\Gamma(m+\beta - \lambda)}{\Gamma(\beta - \lambda)}~
\Biggl[
\frac{1}{\lambda +n+1-\alpha} \cdot \biggl(
\frac{1}{m+\alpha +\beta + \gamma - \lambda -2}
  \nonumber \\
\z  +~
\frac{1}{m+1- \gamma } \biggr)
%\nonumber \\ \z
{}~-~
\frac{1}{\lambda +n+\alpha -1} \cdot \biggl(
\frac{1}{m+ \beta + \gamma - \lambda -1} +
\frac{1}{m+2-\alpha -\gamma } \biggr)\Biggr]  \label{14}
\end{eqnarray}

%Consider the term $\sim (n+ \lambda +1- \alpha )^{-1}$ in the r.h.s. of
%Eq.(\ref{12}). Using Eq.(\ref{C5}) we have
%\begin{eqnarray}
%\z \sum_{n=0}^{\infty}
%\frac{\Gamma(n+2 \lambda )
%%a_n(2\lambda)
%}{n! \Gamma (2 \lambda) }
%\frac{\Gamma (m+\beta +n)}{\Gamma (m+n+1+\lambda)
%%\Gamma (\beta )
%}~
%\frac{1}{\lambda +n+1-\alpha}   \nonumber \\
%\z ~=~\frac{(-1)^m \Gamma(\beta)\Gamma(1-\beta)\Gamma(\lambda +1-\alpha)
%\Gamma(\lambda -1+\alpha)}{\Gamma(m+\alpha)\Gamma(2\lambda)
%\Gamma(2+\lambda-\alpha-\beta-m)}   \nonumber \\
%\z ~-~ \frac{\Gamma(\beta)\Gamma(1-\beta)}{\Gamma(\lambda)\Gamma(1-\lambda)}
%\sum_{n=0}^{\infty}
%\frac{\Gamma(n+2 \lambda )}{n! \Gamma (2 \lambda) }
%\frac{\Gamma (n+\lambda -m)}{ \Gamma (n-m+1+2\lambda-\beta)}~
%\frac{1}{\lambda +n-1+\alpha}  \label{15}
%\end{eqnarray}
%
%\begin{eqnarray}
%\z  \frac{\Gamma(\beta)\Gamma(1-\beta)}{\Gamma(\lambda)\Gamma(1-\lambda)}
% \sum_{m=0}^{\infty}
%\frac{\Gamma (n+\lambda -m)}{\Gamma (n-m+1+2\lambda -\beta)}~
%\frac{\Gamma(m+\beta - \lambda)}{m! \Gamma(\beta - \lambda)}~
%~\frac{1}{m+a}   \nonumber \\
%\z ~=~ \frac{(-1)^n \Gamma(\beta)\Gamma(1-\beta)\Gamma(a)\Gamma(\beta
%  -\lambda -a)}{\Gamma(1-\lambda -n-a)\Gamma(\beta -\lambda)
%\Gamma(1+a+2\lambda-\beta+n)}  \nonumber \\
%\z ~-~  \sum_{m=0}^{\infty}
%\frac{\Gamma (n+m+\beta )}{\Gamma (n+m+1+\lambda )}~
%\frac{\Gamma(m+\beta - \lambda)}{m!\Gamma(\beta - \lambda)}~
%~\frac{1}{m+\beta -\lambda -a}  \label{16}
%\end{eqnarray}

Apply Eq. (\ref{8.2}) with $k=n, ~b=\lambda +1-\alpha $ and with
$k=m, ~b=\alpha + \beta + \gamma - \lambda -2 $ and $b=1-\gamma $
%and (\ref{16}) to the term
%$\sim (n+ \lambda +1- \alpha)^{-1}$
in the r.h.s. of Eq.(\ref{14}).
%subsequantly with
%$a=\alpha+\beta+\gamma-\lambda -2$ and $a=1- \gamma$,
After some algebra the double-sum terms cancel one of another
and we obtain the following form
for the part $\tilde{I}$
%we have
\begin{eqnarray}
\z \tilde{I}~=~
 \frac{ \Gamma(1-\beta)\Gamma(\lambda +1-\alpha)
\Gamma(\lambda -1+  \alpha)\Gamma(1-\beta +\lambda)\Gamma(1- \gamma)
\Gamma(\alpha+\beta+\gamma -\lambda -2)}{\Gamma(2\lambda)
\Gamma(2+\lambda-\alpha -\beta)\Gamma(\alpha +\gamma -1)
\Gamma(2+\lambda-\gamma-\beta)\Gamma(\alpha +\beta -\lambda-1)}
       \nonumber  \\
\z ~-~    \sum_{n=0}^{\infty}
\frac{\Gamma(n+2 \lambda )}{n! \Gamma (2 \lambda) }~
\frac{(-1)^n \Gamma(1-\beta)}{\Gamma(\beta -\lambda)} \cdot
\frac{1}{\lambda +n+\alpha -1}   \label{18}  \\
%[
\z \cdot  \Biggl[
 \frac{\Gamma(\alpha+\beta+\gamma -\lambda -2)
\Gamma(2-\alpha  -\gamma )}{\Gamma(3-\alpha -\beta -\gamma -n)
\Gamma(\alpha +\gamma +\lambda-1+n)} ~+~
%                            \nonumber \\ \z
\frac{\Gamma(1-\gamma)\Gamma(\beta +\gamma  -\lambda -1)}
{\Gamma(\gamma-\lambda -n)
\Gamma(2+2\lambda-\beta-\gamma+n)}
%]
\Biggr]  \nonumber
\end{eqnarray}

Starting from the term $\sim (n+ \lambda -1+ \alpha)^{-1}$ in the r.h.s. of
Eq.(\ref{14}) and repeating the above analysis,
%similar to above one,
we can obtain
 another form of $\tilde{I}$:
\begin{eqnarray}
\z \tilde{I}~=~-~
 \frac{ \Gamma(1-\beta)\Gamma(\lambda +1-\alpha)
\Gamma(\lambda -1+  \alpha)\Gamma(1-\beta +\lambda)\Gamma(2- \alpha -\gamma)
\Gamma(\beta+\gamma -\lambda -1)}{\Gamma(2\lambda)
\Gamma(\lambda+\alpha -\beta)\Gamma(\gamma )
\Gamma(3+\lambda-\alpha -\gamma-\beta)\Gamma(\beta -\alpha-\lambda+1)}
      \nonumber  \\
\z  ~+~   \sum_{n=0}^{\infty}
\frac{\Gamma(n+2 \lambda )}{n! \Gamma (2 \lambda) }~
\frac{(-1)^n \Gamma(1-\beta)}{\Gamma(\beta -\lambda)}  \cdot
\frac{1}{\lambda +n-\alpha +1}    \label{19}  \\
%[
\z  \cdot  \Biggl[
 \frac{\Gamma(\beta+\gamma -\lambda -1)
\Gamma(1-\gamma )}{\Gamma(2 -\beta -\gamma -n)
\Gamma(\gamma +\lambda+n)} ~+~
%                            \nonumber \\ \z
\frac{\Gamma(2-\alpha-\gamma)\Gamma(\alpha+\beta +\gamma  -\lambda -2)}
{\Gamma(\alpha+\gamma-1-\lambda -n)
\Gamma(3+2\lambda-\alpha-\beta-\gamma+n)}
%]
\Biggr]  \nonumber \end{eqnarray}

Applying Eq.(\ref{8.2}) to one of the two terms contained in the sum in
the r.h.s. of
Eqs.(\ref{18})
or (\ref{19}) we can have also two additional representations for $\tilde{I}$:
\begin{eqnarray}
\z \tilde{I}~=~
 \frac{ \Gamma(1-\beta)\Gamma(\lambda +1-\beta )\Gamma(\lambda +1-\alpha)
\Gamma(\lambda -1+  \alpha)\Gamma(1- \gamma)
\Gamma(2-\alpha-\gamma )}{\Gamma(2\lambda)\Gamma(\alpha)
\Gamma(1- \alpha)\Gamma(2+\lambda-\alpha -\beta)
\Gamma(3+\lambda-\alpha -\gamma-\beta)}
       \nonumber  \\
\z  ~+~   \sum_{n=0}^{\infty}
\frac{\Gamma(n+2 \lambda )}{n! \Gamma (2 \lambda) }~
\frac{(-1)^n \Gamma(1-\beta)}{\Gamma(\beta -\lambda)}
\Biggl[
%  \nonumber  \\
%\z   \Bigl[
 \frac{\Gamma(\alpha+\beta+\gamma -\lambda -2)
\Gamma(2-\alpha  -\gamma )}{\Gamma(3+2\lambda-\alpha -\beta -\gamma +n)
\Gamma(\alpha +\gamma -\lambda-1-n)}  \nonumber \\
\z      \cdot   \frac{1}{\lambda +n-\alpha +1}
%                            \nonumber \\ \z
{}~-~ \frac{\Gamma(1-\gamma)\Gamma(\beta +\gamma  -\lambda -1)}
{\Gamma(\gamma-\lambda -n)
\Gamma(2+2\lambda-\beta-\gamma+n)}  \cdot
\frac{1}{\lambda +n+\alpha -1}
%]
\Biggr]  \label{20}
\end{eqnarray}

\begin{eqnarray}
\z \tilde{I}~=~
 \frac{ \Gamma(1-\beta)\Gamma(\lambda +1-\beta )\Gamma(\lambda +1-\alpha)
\Gamma(\lambda -1+  \alpha)\Gamma(\beta+ \gamma -\lambda -1)
\Gamma(\alpha+\gamma+\beta-\lambda-2 )}{\Gamma(2\lambda)\Gamma(\alpha)
\Gamma(1- \alpha)\Gamma(\gamma)
\Gamma(\lambda +\gamma -1)}
       \nonumber  \\
\z  ~+~   \sum_{n=0}^{\infty}
\frac{\Gamma(n+2 \lambda )}{n! \Gamma (2 \lambda) }~
\frac{(-1)^n \Gamma(1-\beta)}{\Gamma(\beta -\lambda)}
\Biggl[
%  \nonumber  \\
%\z   \Bigl[
 \frac{\Gamma(\beta+\gamma -\lambda -1)
\Gamma(1 -\gamma )}{\Gamma(\lambda+\gamma +n)
\Gamma(2-\beta -\gamma -n)}  \cdot
%  \nonumber \\ \z
\frac{1}{\lambda +n-\alpha +1}
                            \nonumber \\ \z   ~-~
\frac{\Gamma(2-\alpha-\gamma)\Gamma(\alpha+\beta +\gamma  -\lambda -2)}
{\Gamma(3-\alpha-\gamma-\beta -n)
\Gamma(\alpha+\lambda+\gamma-1+n)}  \cdot
\frac{1}{\lambda +n+\alpha -1}
%]
\Biggr]  \label{21}
\end{eqnarray}

 Eqs. (\ref{18})-(\ref{21}) are equal to one another because they are
connected by
the transformation (\ref{C5}). For concrete values of $\alpha$, $\beta$ and
$\gamma$ the more convenient one of them may be used.

Thus, a quite simple solution for $A(\alpha, \beta, \gamma)$ is
obtained\footnote{ Before our studies, the possibility to represent
  $C_f[A(\alpha, \beta, \gamma)]$ as a combination of
  $_3F_2$-hypergeometric functions with unit argument, has been
  observed in \cite{Bro}.}. In next section we will consider the
important special case of
 these results.\\

{\bf 4.} {\it As a simple but important example} to apply these
results we consider the diagram $J(1,1,1,1,\alpha )$. It arises in the
framework of a number of calculations (see \cite{Kaz}
%\cite{Gri}, cite{Va}, \cite{Vas} and
- \cite{Gra}). Its coefficient
function $I(\alpha )
\equiv C_f[J(1,1,1,1,\alpha )]$ can be found as follows
\begin{eqnarray} \z
I(\alpha )~ \stackrel{f}{=}
\frac{a_0^4(1)a_0(\alpha )}{a_0(\alpha +2-2\lambda)}~
C_f[J( \lambda,\lambda,\lambda,\lambda,\tilde \alpha )] \mbox{ and }
C_f[J( \lambda,\lambda,\lambda,\lambda,\tilde \alpha )] =
C_f[A(\tilde \alpha ,3-\lambda -\tilde \alpha , \lambda)]
% \nonumber\\ \z
\nonumber
\end{eqnarray}
The latter equation may be obtained by analogy with (\ref{10}).

{}From Eqs. (\ref{13}) and (\ref{18}) we obtain
   \begin{eqnarray}
\z I(\alpha )~=~ - \frac{2}{(4\pi)^D}
 \frac{ \Gamma ^2(\lambda)\Gamma(\lambda - \alpha )
\Gamma(\alpha +1 -2 \lambda )}{\Gamma(2\lambda)
\Gamma(3\lambda -\alpha  -1)}
       \label{4.1}  \\
\z  \cdot \Biggl[
\frac{ \Gamma ^2(1/2)\Gamma(3\lambda - \alpha -1)\Gamma(2\lambda - \alpha )
\Gamma(\alpha +1 -2\lambda )}{\Gamma(\lambda)
\Gamma(2\lambda +1/2 -\alpha )\Gamma(1/2-2\lambda +\alpha )}
{}~+~   \sum_{n=0}^{\infty}
\frac{\Gamma(n+2 \lambda )}{ \Gamma (n+\alpha +1) }~
\frac{1}{n+1 -\lambda +\alpha }
\Biggl]   \nonumber
\end{eqnarray}
Note that in \cite{Kaz} Kazakov has got another result for $I(\alpha )$:
   \begin{eqnarray}
\z I(\alpha )~=~ - \frac{2}{(4\pi)^D}
 \frac{ \Gamma ^2(\lambda)\Gamma(1-\lambda)\Gamma(\lambda - \alpha )
\Gamma(\alpha +1 -2 \lambda )}{\Gamma(2\lambda)\Gamma(\alpha )
\Gamma(3\lambda -\alpha  -1)}
       \label{4.2}  \\
\z \cdot \Biggl[
\frac{ \Gamma(\lambda)\Gamma(2- \lambda)\Gamma(\alpha )
\Gamma(3\lambda -\alpha  -1)}{
\Gamma(2\lambda -1)\Gamma(3-2\lambda )}
{}~-~   \sum_{n=0}^{\infty} (-)^{n}
\frac{\Gamma(n+2 \lambda )}{ \Gamma (n+2-\lambda) }~ \biggl(
\frac{1}{n+1 -\lambda +\alpha } ~+~ \frac{1}{n+2\lambda -\alpha } \biggr)
\Biggl]   \nonumber
\end{eqnarray}

{}From Eqs. (\ref{4.1}) and (\ref{4.2}) we obtain the
transformation rule for $_3F_2$-hypergeometric function with argument $-1$:
\begin{eqnarray}
\z
_3F_2(2a,b,1;b+1,2-a;-1)~=~b \cdot
\frac{\Gamma(2-a)\Gamma(b+a-1)\Gamma(b-a)
\Gamma(1+a-b)}
{\Gamma(2a)\Gamma(1+b-2a)}
\label{4.3}  \\
\z -~ \frac{1-a}{b+a-1} \cdot {}_3F_2(2a,b,1;b+1,b+a;1)
-~ \frac{b}{1+a-b} \cdot {}_3F_2(2a,1+a-b,1;2+a-b,2-a;-1),
  \nonumber
\end{eqnarray}
where $a=\lambda$ and $b=1-\lambda + \alpha $ are used.

Equation (\ref{4.3}) has been explicitly checked at $a=1$ and $b=2-a$
(i.e. $\lambda =1$ and $\alpha =1$), where the
$_3F_2$-hypergeometric functions may be calculated
exactly. We cannot directly prove Eq.(\ref{4.3}) at arbitrary $a$ and $b$
values: the general proof seems to be non-trivial.
Note that it is different from the equations of \cite{PBM} and
may be considered
as a new transformation rule.\\

{\bf 5.} {\it As a conclusion} we note that we obtained the results for
the class of master two-loop diagrams containing the vertex with two
propagators, having indices $1$ or $\lambda = D/2 -1$.
As a by-product of the above-mentioned studies, the new transformation
rule  of
$_3F_2$-hypergeometric series with argument $-1$, was obtained. It
may be used together with other rules (see Ref.\cite{PBM}, for example).

We would like to note also, that there is only one other diagram
$$B(\alpha, \beta, \gamma ) ~=~ J(\alpha,\lambda, \beta, \lambda,
\gamma)$$
with two indices $\lambda$, which does not fall into the above-considered FD
class. The diagram
$ J(\tilde \alpha,\lambda,\tilde \beta, \lambda,\tilde \gamma)$
can be obtained from $B(\alpha, \beta, \gamma)$ by a analysis similar
to (\ref{9}) and
it cannot be considered as the independent one.
The incorporation of the diagram $B(\alpha, \beta, \gamma)$ in our
analysis invites futher investigations and it is
the subject of the next publication \cite{4}.\\

%\section{Acknowledgments} \indent

  The author is grateful to Prof. A.N.Vasil'ev for a stimulation of this
 research and Prof. D.Broadhurst for information about a
 possibility to represent FD
%Feynman diagram
  $A(\alpha, \beta, \gamma)$ as a combination of
  $_3F_2$-hypergeometric functions with unit argument (see  \cite{Bro}).

 The author is grateful also to Prof. P.Aurenche, Prof. D.Broadhurst,
Prof. K.G.Chetyrkin, Dr. A.L.Kataev, Prof. D.I. Kazakov,
Dr. N.A.Kivel, Dr. A.S. Stepanenko and Prof. A.N.Vasil'ev  for discussions.

\setcounter{secnumdepth}{2}
\addcontentsline{toc}{section}{APPENDIX}

\setcounter{section}{0}
\setcounter{subsection}{0}
\setcounter{equation}{0}
 \def\thesection{\Alph{section}}
\def\thesubsection {\thesection.\arabic{subsection}}
\def\theequation{\thesection.\arabic{equation}}

%\appendix
\section{Appendix
%A
}

Here we present useful formulae to use of Gegenbauer polynomials.
% and traceless products.
Following \cite{2,2.1}, $D$-space integration can
be represented in the form
$$
d^Dx~=~ \frac{1}{2} S_{D-1}(x^2)^{\lambda} dx^2 d\hat{x},
$$
where $
%\bf{x}^2=\vec{x}^2,
{}~\hat{x}=\vec{x}/\sqrt{
%\vec{x}
x^2}$, and
$S_{D-1}=2 {\pi}^{\lambda+1}/\Gamma(\lambda+1)$ is the surface of the
unit hypersphere in $R_D$. The Gegenbauer polynomials
$C_n^{\delta}(t)$ are defined as \cite{1.1,2.2}
\begin{eqnarray}
 (1-2rt + r^2)^{-\delta} ~=~ \sum^{\infty}_{n=0}~C_n^{\delta}(t)r^n
{}~~~~(r \leq 1),~~~~ C_{n}^{\delta}(1) ~=~ \frac{ \Gamma(n+2\delta)}
{n!\Gamma(2\delta)}
  \label{A1}
\end{eqnarray}
whence the expansion for the propagator is:
\begin{eqnarray}
\frac{1}{(x_1-x_2)^{2\delta}} ~=~
\sum^{\infty}_{n=0} ~C_n^{\delta}(\hat{x_1}\hat{x_2})~
\Biggl[
\frac{{(x^2_1)}^{n/2}}{{(x^2_2)}^{n/2 + \delta}} \Theta(x_2^2 - x_1^2)
{}~+~
\Bigl(x_1^2 \longleftrightarrow x_2^2  \Bigr)  \Biggr],
  \label{A2}
\end{eqnarray}
where
\[ \Theta(y)~=~ \left\{
\begin{array}{rl}
1, & \mbox{ if }y \geq 0 \\
0, & \mbox{ if }y  <   0
\end{array} \right.  \]

Orthogonality of the Gegenbauer polynomials is expressed by the
equation (see \cite{2})
\begin{eqnarray}
\int ~C_n^{\lambda}(\hat{x_1}\hat{x_2})~
 \hat{x_2}
 ~C_m^{\lambda}(\hat{x_2}\hat{x_3})~=~\frac{\lambda}{n+\lambda} \delta^m_n
 ~C_n^{\lambda}(\hat{x_1}\hat{x_3}),~
  \label{A3}
\end{eqnarray}
where $\delta^m_n$ is the Kroneker symbol.

%To our study
The following formulae are useful (see \cite{2,3}):
\begin{eqnarray}
\z
C_n^{\delta}(x)~=~
\sum_{p \geq 0} \frac{(2x)^{n-2p}(-1)^p \Gamma(n-p+\delta)}
{(n-2p)!p!\Gamma(\delta)}
%\label{A4}   \\\z
{}~~~
{}~ \mbox{ and }~
\nonumber   \\ \z
\frac{(2x)^{n}}{n!} ~=~\sum_{p \geq 0}
C_{n-2p}^{\delta}(x) \frac{(n-2p+\delta)\Gamma(\delta)}
{p!\Gamma(n-p+\delta+1)}  \label{A4}
%\\ \z   C_{n}^{\delta}(1) ~=~ \frac{ \Gamma(n+2\delta)}
%{n!\Gamma(2\delta)}  \label{A6}
\end{eqnarray}

Substituding the latter equation from (\ref{A4}) for $\delta = \lambda$ to
the first one, we have the following equation after the separate
analysis at odd and even $n$:
\begin{eqnarray}
C_n^{\delta}(x)~=~\sum_{k=0}^{[n/2]}
C_{n-2p}^{\lambda}(x) \frac{(n-2k+\lambda) \Gamma(\lambda)}
{k!\Gamma(\delta)}  \frac{ \Gamma(n+\delta-k)\Gamma(k+\delta-\lambda)}
{\Gamma(n+\lambda+1-k)\Gamma(\delta-\lambda)}  \label{A7}
\end{eqnarray}

This equation is an example of the
%one from
%basical
%more used
equations used in our analysis.
Note that alternatively to (\ref{A7}) we can use  the expansion of
$(x_1-x_2)^{-2\delta}$ similar to (\ref{A3}) but directly in a series
of GP $C_n^{\lambda}(\hat x_1 \cdot \hat x_2)$ (see \cite{2.1}).

\end{document}